\documentclass{aastex}
\usepackage{emulateapj}
\usepackage{onecolfloat}
\usepackage{natbib}
\bibpunct{(}{)}{;}{a}{,}{,}

\input epsf

\def\plottwo#1#2{\centering \leavevmode
    \epsfxsize=1.0\columnwidth \epsfbox{#1} \hfil
    \epsfxsize=1.0\columnwidth \epsfbox{#2}}


\long\def\comment#1{}

\def\W2{{\cal W}}

\def\phz{photo-{\it z}}
\def\be{\begin{equation}}
\def\ee{\end{equation}}
\def\bea{\begin{eqnarray}}
\def\eea{\end{eqnarray}}

\def\cmm2{{\,\rm cm^{-2}}}
\def\cm2{{\,{\rm cm}^2}}
\def\cmm3{{\,{\rm cm}^{-3}}}
\def\gcmm3{{\,{\rm g\,cm^{-3}}}}

\def\fun#1#2{\lower3.6pt\vbox{\baselineskip0pt\lineskip.9pt
  \ialign{$\mathsurround=0pt#1\hfil##\hfil$\crcr#2\crcr\sim\crcr}}}

\lefthead{Knox et al.} \righthead{Weighing the Universe}

\begin{document}
%%% \nopagebreak
%%% \vspace{-\baselineskip}
\bibliographystyle{apj}
\twocolumn[%%% Begin front material
\submitted{To be submitted to ApJ}
\title{Weighing the Universe with Photometric
Redshift Surveys and the Impact on Dark Energy Forecasts}
\author{Lloyd Knox$^1$, Yong-Seon Song$^2$ and Hu Zhan$^1$}
\affil{$^{1}$ Department of Physics, University of California,
Davis, CA 95616, USA, email:
lknox@ucdavis.edu,zhan@bubba.ucdavis.edu}
\affil{$^2$ Department of Astronomy and Astrophysics, University of
Chicago, 5640 South Ellis Avenue, Chicago, Illinois 60637}

\begin{abstract}
With a wariness of Occam's razor awakened by the discovery of cosmic
acceleration, we abandon the usual assumption of zero mean curvature
and ask how well it can be determined by planned surveys.  We also
explore the impact of uncertain mean curvature on forecasts for the
performance of planned dark energy probes.  We find that weak lensing
and photometric baryon acoustic oscillation data, in combination with
CMB data, can determine the mean curvature well enough that the
residual uncertainty does not degrade constraints on dark energy.  We
also find that determinations of curvature are highly tolerant of
photometric redshift errors. 
\end{abstract}

 ]%%% End front material

\section{Introduction}

Due to indications from the CMB that the mean spatial curvature
is close to zero, and the empirical successes of inflation, it has
become quite common to assume that the mean curvature is exactly zero
in analyses of current data \citep[e.g.][]{spergel03}) and in
forecasting cosmological constraints to come from future surveys
\citep[e.g.][]{song04}).  Recently, however there has been renewed
interest in the possibility of non-zero mean spatial curvature.  For
example, \citet{linder05d} explored the impact of dropping the
flatness assumption on the ability of supernova + CMB data to
determine dark energy parameters.  \citet{knox06} quantified how the
combination of distance measurements into the dark energy dominated
era, combined with CMB observations, could be used to determine the
mean spatial curvature.  \citet{bernstein05} considered purely
geometrical constraints on curvature to come from weak lensing (WL)
and baryon acoustic oscillation (BAO) data.  In this paper
we extend Linder's work to other cosmological probes (WL and BAO) and
extend that of \citet{knox06} by forecasting constraints on curvature
to come from specific surveys rather than idealized measurements to
single distances.

This renewed interest in mean curvature is due to several factors.
First, the discovery of cosmic acceleration has made us wary of
Occam's razor, the idea that the simplest possible outcome is the most
likely.  Occam's razor, before data convinced us otherwise, pointed
toward a flat Universe with zero cosmological constant.  Although zero
mean spatial curvature is still consistent with the data, small
departures are also still allowed.  Given how Occam's razor has misled
us in the past, we trust the argument for simplicity less and lend
more credence to the possibility of small, but non-zero, mean spatial
curvature.

Second, recent theoretical work suggests that detectable
amounts of mean spatial curvature from inflation may not be entirely
improbable.  \citet{freivogel06} estimate the probability distribution
of $\Omega_{\rm tot}$ that follows from specific assumptions about the
distribution of the shapes of potentials in the string-theory
landscape.  Taking $N=62$ as a lower bound on the number of e-foldings
(in order to get $\Omega_{tot} > 0.98$, their interpretation of the
current lower bound) they find 10\% of the probability in the range
$62 < N < 64$ or roughly $0.02 > 1-\Omega_{\rm tot} > 4 \times
10^{-4}$.  At face value this says that if we achieve the sensitivity
to mean curvature possible with Planck and high-precision BAO
measurements, as forecasted in \citet{knox06}, there will
be a ~10\% chance of making a detection.  Note, however, that possible
volume factors in the measure, that could strongly favor Universes
undergoing longer periods of inflation, were (knowingly) neglected.  The
importance of these measures is a controversial topic
\citep[e.g.][]{linde95,garriga99}.

Related to these two reasons is a third deriving from the importance
for fundamental physics of a discovery that the dark energy is not a
cosmological constant.  The implications of such a discovery would be
sufficiently dramatic that all the assumptions underlying it would
need to be revisited.  We risk making the following type of error:
claiming detection of non-cosmological constant dark energy, when the
data are actually explained by a cosmological constant plus curvature.
Evidence for non-cosmological constant dark energy would stimulate the
revisiting of many assumptions of the standard cosmological model,
such as the adiabaticity of the primordial fluctuations
\citep[][]{trotta03,trotta04}.

We are {\em not} declaring that dark energy
conclusions reached by assuming $\Omega_{\rm K} = 0$ are uninteresting.  A
result that informed us we need either non-$\Lambda$ dark energy or
non-zero mean curvature would be frustratingly ambiguous, but
nonetheless terribly interesting.

Here we quantify how well planned surveys can measure the mean 
spatial curvature as well as the impact of dropping the flatness 
assumption on the expected dark energy constraints.  In
section ~\ref{sec:surveys} we describe our modeling of the surveys
considered, including systematic errors from, e.g., supernova mean
absolute magnitude evolution and photometric redshift errors.  In
section ~\ref{sec:curvature} we show the constraints on mean curvature
from these surveys individually and in combinations.  In section
~\ref{sec:darkenergy} we do the same for dark energy. Finally, in Section
~\ref{sec:discuss} we discuss and conclude.

As an historical aside we note that our title alludes to an earlier paper 
``Weighing the Universe with the CMB'' \citep{jungman96b} that
pointed out the sensitivity of the location of the CMB acoustic peak
to the mean curvature, and thus the mean density (in units of the
critical density).  Since then CMB observations have indeed been
used to greatly improve the precision with which the mean
curvature is known \citep{miller99,dodelson00,debernardis00}
and most recently \citet{spergel06}.  Further improvements in
precision require determinations of distances to redshifts much
lower than that of the last-scattering surface \citep{eisenstein05,knox06}
made possible by the surveys we discuss here.  

\section{Surveys}
\label{sec:surveys}

The three probes we consider here are probes of the dark energy via
the distance-redshift relation, $D(z)$.  Weak lensing is also
sensitive to dark energy via its influence on the growth of
large-scale structure.  We first emphasize their
qualitative distinguishing characteristics before moving on to a
description of the specific surveys we consider and how we model them.

Supernovae (SN) determine the shape of the distance-redshift curve,
but not an overall amplitude\footnote{The amplitude of $D(z)$ can be
pinned down by complementary determinations of the distance-redshift
relation at low redshift, where to first order in $z$, $D(z) =
H_0^{-1} z$ depends only on one parameter.}.  Planned space-based
supernova surveys probe this relation out to $z \simeq 1.7$
\citep[e.g.][]{aldering05}.  Toward higher redshifts spectral features
used to type the supernovae are at wavelengths to which the detectors
are not sensitive.  The detectors are transparent at these wavelengths
by design, to reduce thermal noise.  Although JWST\footnote{James Webb
Space Telescope: http://www.jwst.nasa.gov} will be capable of
detecting supernovae at higher redshifts, one can expect greater
evolutionary effects at these redshifts.  Indeed, \citet{riess06}
recently pointed out that JWST could be used to study evolutionary
effects at $z \simeq 2$ so that they can be better understood at lower
redshifts.  Also, gravitational lensing contributes to the luminosity
dispersion and this contribution increases with redshift \citep{holz98}.

Baryon acoustic oscillations (BAO) also determine the shape of the $D(z)$
curve, by exploiting a standard ruler in the galaxy power spectrum.
The length of this ruler, the sound horizon at the epoch of CMB last
scattering, can be accurately determined from CMB observations, thus
giving us the amplitude of the $D(z)$ relation as well.  This technique
can be used to redshifts of 3 and beyond.

Weak lensing (WL) observations constrain the $D(z)$ curve in a less
direct manner.  For forecasts of $D(z)$ and growth factor $g(z)$
reconstructions from WL data see \citet{knox05b}.  For an excellent 
discussion of the origin of the $D(z)$ constraints from WL
see \citet{zhang05}.  Like BAO, WL can be used to study $D(z)$
to redshifts of 3 and beyond.  Unlike SN and BAO, the WL technique is
sensitive to the growth of structure, which can also contribute
significantly to the constraints on dark energy \citep{zhang05}.

The SN technique can achieve strong constraints on $D(z)$ with a
sufficiently small number of sufficiently bright objects that a survey
can be designed that allows for spectroscopic redshifts to be
determined for all the objects.  In contrast, weak lensing
observations rely on the shape determinations of very large numbers of
very faint galaxies, making spectroscopy prohibitively
expensive\footnote{Observations of redshifted 21cm lines may make
spectroscopic WL possible eventually.}.  Instead, one must rely on
photometrically-determined redshifts.  The demands on control of
systematic errors on these redshifts are quite stringent
\citep{bernstein04,ma06,huterer05}.  If spectroscopy is used then the
BAO technique requires fewer galaxies than are required by WL;
interesting constraints are possible from ambitious, but achievable,
spectroscopic surveys \citep[e.g.][]{seo03}.  Here we only consider
BAO surveys that forego spectroscopy and thus rely on photometric
redshifts.  We have previously argued that photometric BAO surveys
also place stringent demands on the level of required systematic error
control \citep{zhan06c}.  Although, if one gives up the radial
information and uses tomographic galaxy angular power spectra, the
demands can be greatly reduced (Zhan 2006, in preparation).

We allow the following cosmological parameters to vary in our
forecasts: the dark energy equation-of-state parameters $w_0$ and
$w_a$ as defined by $w(a) = w_0 + w_a(1-a)$, the matter density
$\omega_{\rm m}$, the baryon density $\omega_{\rm b}$, the angular
size of the sound horizon at the last scattering surface $\theta_{\rm
s}$, the equivalent matter fraction of curvature $\Omega_{\rm K}$, the
optical depth to scattering by electrons in the reionized
inter-galactic medium, $\tau$, the primordial helium mass fraction
$Y_{\rm p}$, the spectral index $n_{\rm s}$ of the primordial scalar
perturbation power spectrum, the running of the spectral index
$\alpha$, and the normalization of the primordial curvature power
spectrum $\Delta_R^2$ at $k = 0.05\,\mbox{Mpc}^{-1}$. The fiducial
model has $(w_0, w_a, \omega_{\rm m}, \omega_{\rm b}, \theta_{\rm s},
\Omega_{\rm K}, \tau, Y_{\rm p}, n_{\rm s}, \alpha, \Delta_R^2) = (-1,
0, 0.127, 0.0223, 0.596\,\deg, 0, 0.09, 0.24, 0.951, 0, 2 \times
10^{-9})$. This model is consistent
with the 3-year {\it WMAP} data \citep{spergel06} and has a 
reduced Hubble constant of $h = 0.73$.

\subsection{Supernovae}

Our fiducial supernova data set has 3000 supernovae from a space-based
survey distributed in redshift uniformly from $z=0.4$ to $z=1.7$, 700
supernovae from a ground-based survey distributed uniformly in redshift
from $z=0.2$ to $z=0.7$ and 500 supernovae from a local sample 
distributed from $z=0.02$ to $z=0.1$.  We model the supernova 
effective apparent magnitudes, after standardization 
(e.g., by exploitation of the Phillips relation \citep{phillips93}), as
\be
m_i = M + \alpha_1 z_i + \alpha_2 z_i^2 + 5\log_{10}(D_L(z_i)/10{\rm pc}) + n_i
\ee
where $M$ is the (unknown) absolute magnitude of a $z=0$ 
standardized supernova, the
$z$ and $z^2$ terms allow for a drift in this mean due to evolution effects
or other systematic errors and $n_i$ is from any random sources of scatter,
either intrinsic to the supernovae or measurement noise.  

We assume $\langle n_i n_j \rangle = \sigma^2 \delta_{ij}$ with 
$\sigma = 0.14$ and further that we are able to constrain the
systematic error terms well enough to place Gaussian priors on their
distributions with standard deviations 
$\sigma_P(\alpha_1) = \sigma_P(\alpha_2) = 0.015$.

\subsection{Weak Lensing}

Our fiducial WL survey is modeled after LSST\footnote{Large Synoptic Survey
Telescope:  http://www.lsst.org}.  We use the
source-redshift distribution $n(z) = n_0 z^2 e^{-z/0.5}$ with $n_0$
chosen so the total source density is 50 galaxies per
sq. arcmin.  We assume the shape noise variance increases with redshift as
$\gamma_{\rm rms}(z) = 0.18 + 0.042z$.  We take $40 \le \ell \le 2000$
and a sky coverage of 20,000 square degrees.

We divide the galaxies into nine  photometric redshift (\phz) bins 
evenly spaced from $z_{\rm p} = 0$ to $3.5$, where the subscript $p$ 
distinguishes \phz{}s from true redshifts. Uncertainties in the 
error distribution of \phz{}s are treated as in 
\citet{ma06}. Specifically, we define an rms \phz{} standard
deviation $\sigma_z$
and a \phz{} bias $\delta z$ at each of 35 redshift values
evenly spaced over the range $z_{\rm p} = 0$ to $3.5$.  
The \phz{} bias and rms at an arbitrary redshift are 
linearly interpolated from the 70 parameters. This treatment of \phz{}
uncertainties is based on our expectation that \phz{} calibrations 
through spectroscopy and other means (e.g. Schneider et al. 2006, in
preparation; Newman 2006, in preparation) will be available, though 
challenging, at redshift intervals of width $\sim 0.1$.

We adopt a conservative estimate of the rms \phz{} error of
$\sigma_z=0.06(1+z)$ with $\delta z = 0$.
Smaller dispersions have been achieved for the CFHT Legacy Survey for
a sample of galaxies with $i'_{\rm AB}$ magnitudes less than
24 and $z \la 2$\citep{ilbert06} .  We assume that through a
calibration process we will know the \phz{} bias parameters to within
$\pm \sigma_P(\delta z_i)$ which in the following ranges from an
optimistic 0.001 to a pessimistic 0.01.  The prior we assume for the
rms parameter takes on a similar range.  To reduce the dimensions of
the parameter space we explore, we always set $\sigma_P(\sigma_{zi}) =
\sqrt{2}\sigma_P(\delta z_i)$.

\subsection{Baryon Acoustic Oscillations} 

Our BAO survey uses the same galaxies and \phz{} parameters as the WL
survey. We only include the angular power spectrum in our forecast, 
not the redshift-space power spectrum,
because to extract information from the radial clustering one has to 
meet very stringent \phz{} requirements \citep{zhan06c}.

Unlike the WL shear power spectrum, the galaxy angular power spectrum 
has a narrow kernel, which is the radial galaxy distribution in 
the true-redshift space. Hence, one can use more \phz{} bins for BAO
until shot noise overwhelms the signal or the bin sizes are much 
smaller than the rms \phz{} errors (Zhan, in preparation). 
For this work, we divide the galaxies into 30 \phz{} bins from 
$z_{\rm p} = 0.15$ to $3.5$ with bin size proportional to $1+z$. 

To avoid contamination by nonlinearity, we exclude modes that have
the dimensionless power spectrum $\Delta^2(k) > 0.4$, e.g., 
$k_{\rm max} \sim 0.15\,h\,\mbox{Mpc}^{-1}$ or $\ell_{\rm max} = 68$ 
at $z = 0.15$. We then only use multipoles $40 \le \ell \le 3000$ for 
BAO.

We treat the galaxy clustering bias in the same way as the \phz{} 
parameters, i.e., we assign 35 bias parameters $b_i$ uniformly 
from $z = 0$ to $3.5$ and linearly interpolate the values. The fiducial
bias model is $b = 1 + 0.84 z$, and we apply a prior of 20\% to each 
bias parameter. 

We implement the method by \citet{hu04b} to combine BAO and WL. 
Note, however, that we do not use the halo model to calculate the
galaxy bias, since we restrict our analysis to largely linear 
scales.  

\subsection{CMB and $H_0$} 

We calculate a Fisher matrix from the CMB data expected from
\emph{Planck}, following the treatment in \citep{kaplinghat03b}.
We also
create a Fisher matrix for the HST Key Project Hubble constant 
determination \citep{freedman01} by projecting
a Hubble constant constraint of $\sigma_{\rm P}[\ln H_0] = 0.11$ 
into our parameter space.   We add both these Fisher matrices
to all of the Fisher matrices calculated for the dark energy
probes.  Of course, when plotting combinations we are careful
to only add these Fisher matrices in once to avoid double-counting
the $H_0$ and CMB information.  

\subsection{Systematic Errors}

The surveys we consider, if they are to achieve the errors we forecast
below, will need to achieve exquisite control of systematic errors.  Our
data modeling includes photometric redshift errors and supernova evolution
but not many other sources of systematic error.  Even for these that
we do include, our modeling may not be sufficiently general to adequately
model the real world.  We need more data to know for sure.  

For further discussion of systematic errors from a space-based SN
mission see \citet{kim04}.  For WL we have not included galaxy shape
measurement systematics \citep[see e.g.][]{huterer06} and intrinsic
alignments, most importantly alignments between source galaxies and
shear \citep{hirata04,mandelbaum06}.  For further discussion of these
effects, see the technical appendix of the impending report of the
Dark Energy Task Force (DETF).  Also note that recent work with
archival Subaru data demonstrates that very low levels of spurious
additive shear are achievable from the ground \citep{wittman05}.  For
BAO, perhaps the major concern is spatially varying photometric
offsets.  Controlling photometry offsets at the requisite levels was a
significant challenge for the recent BAO analysis from SDSS
photometric data \citep{padmanabhan06}.  Spatially variable dust
extinction must be controlled as well, as discussed in
\citet{zhan06d}.

\section{Curvature}
\label{sec:curvature}

\begin{figure}[b]
\centerline{\scalebox{.4}{\includegraphics{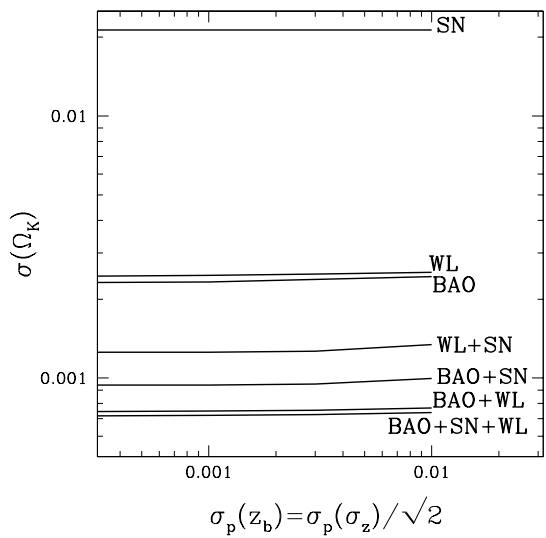}}}
\caption{Redshift error priors vs. $\sigma(\Omega_{\rm K})$ for various 
data sets and combinations.  
}
\label{fig:curv}
\end{figure}

Results for curvature are shown in Fig.~\ref{fig:curv}.  The
first striking feature is the robustness to the quality of
the photometric redshifts.  This behavior is in contrast
to that of the dark energy constraints from weak lensing as
calculated in \citet{bernstein04,ma06,huterer06} and in
the next section.

For the supernova curve this robustness is trivial; we assume a
spectroscopic survey for the supernovae and hence the results, by
design, are completely independent of the photometric redshift
parameters.  For WL and BAO the near lack of dependence has its
origins in the critical role played by measurements of distances to
redshifts in the matter-dominated era \citep{knox06}.  The comoving
angular-diameter distance varies very slowly with $z$ at 
$z \ga 2$, and hence the tolerance to redshift errors is quite high.

The value of $\sigma(\Omega_{\rm K})$ achieved depends on which probe
is used.  Those that reach to larger redshifts (WL and BAO) are
protected from the confusing effects of dark energy and thus do a
better job of determining $\Omega_{\rm K}$.  WL and BAO data sets,
either individually or in combination, can be used to determine
$\Omega_{\rm K}$ at about the $10^{-3}$ level, consistent with the
much more model-independent estimates in \citet{knox06}.

In addition to the shorter redshift reach, another important difference
of the SN probe is the lack of strong normalization of $D(z)$.  
This can be remedied by a better measurement of the Hubble constant.
Improving the Hubble constant prior to 1\% reduces the supernova
$\sigma(\Omega_{\rm K})$ to 0.004.  

These forecasts for curvature are correct if the dark energy
is parameterized by $w_0$, $w_a$ and $w_0+1 \simeq w_a \simeq 0$.
If the dark energy density is more important at higher redshifts
than in our fiducial model, then the constraints will weaken
somewhat.  

If the history of the dark energy equation-of-state parameter 
is not well-approximated by our assumed form, and has a density
higher at $z \ga 3$ than its current value, then we risk
significant systematic errors in our determination of $\Omega_{\rm K}$.  
We can partially guard against this error by verifying
that $w_0$ and $w_a$ can be adjusted to give a good fit to the
data.  The worry remains though that unexpectedly high dark
energy density at $3 < z < 1100$ could mimic a small negative 
curvature.  

How much model dependence will obscure our determination of the
curvature depends on how the experimental program plays out.  One of
the most interesting possible results is that $K$ is determined to be
greater than zero ($\Omega_{\rm K} < 0$) with high confidence.  Such a
result would be difficult to reconcile with inflation because short
inflation scenarios solve the horizon problem by bubble nucleation
which leads to $K > 0$.  A determination that $K > 0$ would be
challenging to the whole string theory landscape paradigm.  Further,
it would be robust to the systematic error described above, since
unexpectedly high dark energy density, if unaccounted for, would mean
the true value of $K$ is even larger.

\citet{bernstein05} pointed out that gravitational lensing's
sensitivity to the source-lens angular-diameter distance 
means that the curvature could be determined in a manner independent
of assumptions about $H(z)$.  Bernstein's effect contributes
to the curvature constraints we forecast, but at a highly
subdominant level.  Were we to allow more freedom in the possible
time variations of the dark energy density Bernstein's 
effect would become important.  Its independence of Einstein's
equations is an interesting virtue.  

\section{Dark Energy}
\label{sec:darkenergy}

\begin{figure*}
\plottwo{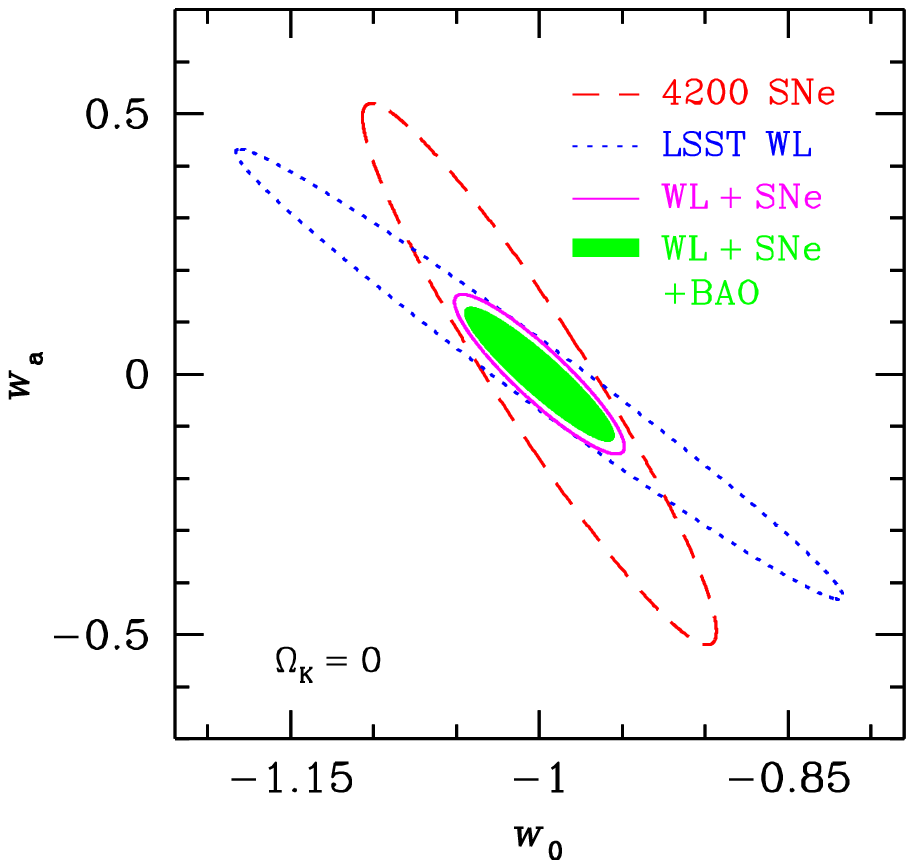}{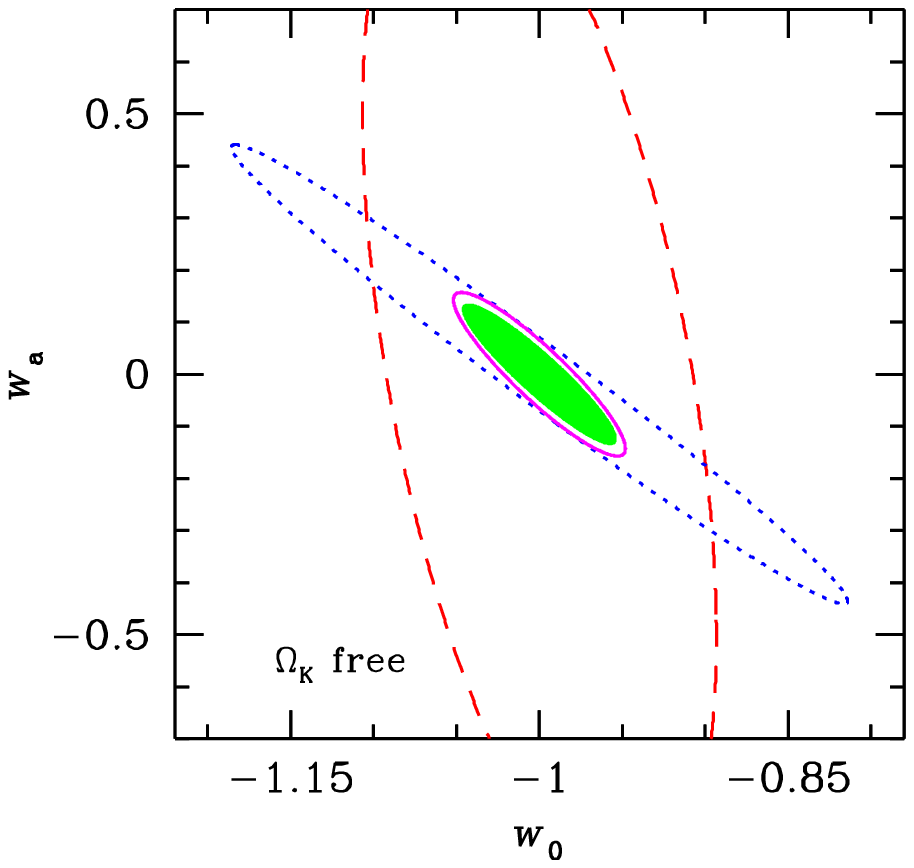}
\caption{\emph{Left panel}: 1,\ 68\% confidencce error contours of $w_0$ and 
$w_a$ assuming a flat universe. The contours are shown for 4200 SNe 
(dashed line), LSST WL (dotted line), the combination WL + SN 
(solid line), and the combination WL + SN + BAO (shaded area).
All results include 
\emph{Planck} and $H_0$ priors and assume $\sigma_{\rm P}(\delta z) = 
\sigma_{\rm P}(\sigma_z) / \sqrt{2} = 0.003$. \emph{Right panel}:
the same as the left panel but with $\Omega_{\rm K}$ treated as a free
parameter.}
\label{fig:w0wa}
\end{figure*}

The SN, WL, and BAO constraints on $w_0$ and $w_a$ are 
given in Figure~\ref{fig:w0wa} with (left panel) and without 
(right panel) assuming a flat universe. The SN constraint 
(dashed lines) on $w_a$ is very sensitive to curvature 
\citep[as shown by][]{linder05d}, whereas WL (dotted lines), being
able to determine the curvature parameter $\Omega_{\rm K}$ to
$\sim 0.001$ (see Figure~\ref{fig:curv}), is not. 
Hence their combination (solid lines) is only slightly 
affected by adding $\Omega_{\rm K}$ as a free parameter. 

The degeneracy between $\Omega_{\rm K}$ and $w(a)$ given SN data can
be understood as follows.  The SN data (with no CMB or $H_0$ data
added) are only sensitive to $\Omega_m$, $\Omega_{\rm K}$ and $w(a)$.
Including the constraint from the CMB on the distance to last
scattering can be thought of as pinning down one combination of
$\Omega_m$ and $\Omega_K$.  The remaining degree of freedom has some
degeneracy with $w(a)$ and is what is responsible for degrading the
$w(a)$ constraints.  Including an $\Omega_m$ prior would provide
the one extra constraint necessary to remove the
degeneracy.  As \citet{linder05d} showed, the results with a prior
$\sigma(\Omega_m)=0.01$ are very similar to the results with curvature
fixed.  Similar improvements would come from a strong Hubble constant
prior since, combined with Planck's determination of $\Omega_m h^2$ to
better than 1\%, this can be translated into a determination of
$\Omega_m$.  The importance of $H_0$ for dark energy probes has been
stressed by \citet{hu05}.  

\begin{figure}[b]
\centerline{\scalebox{.4}{\includegraphics{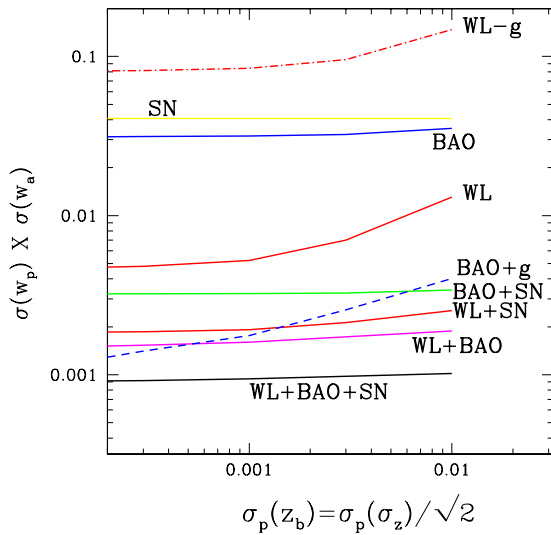}}}
\caption{Redshift error priors vs. the dark energy equation-of-state 
parameter error product $\sigma(w_p) \times \sigma(w_a)$
for various data sets and combinations.
See text for explanations of BAO+g and WL-g.    
 }
\label{fig:fom}
\end{figure}

We now turn to the dependence of our forecasted dark energy
constraints on photometric redshift errors.  We show our results in
Figure~\ref{fig:fom}.  Specifically, we plot $\sigma(w_p) \times
\sigma(w_a)$ as we change the priors on the mean redshift of each
redshift bin and the rms of the scatter in each redshift bin.
\citet{hu03a} introduced the variable $w_p = w(a_p)$ where $a_p$ is
the scale factor at which $w(a)$ is determined with the smallest
uncertainty, assuming that $w(a) = w_0 + (1-a) w_a$.  From this
definition it also follows that the errors on $w_p$ and $w_a$ are
uncorrelated with each other.  The product $\sigma(w_p) \times
\sigma(w_a)$ is proportional to the area of the $w_0$, $w_a$ 95\%
confidence ellipse.  The inverse of this product is proportional to
the figure of merit used by the DETF, recently
discussed by \citet{martin06} and \citet{linder06}.

Just as in Figure~\ref{fig:curv} the SN alone case is 
a horizontal line because the photo-z parameters have nothing
to do with the SN data.  We also see the dependence
of $\sigma(w_p) \times \sigma(w_a)$ for WL on the photometric redshift
parameters.  For these forecasts we have marginalized over all
of our other parameters, including curvature.  With curvature
fixed the SN alone case would improve from 
$\sigma(w_p)\times \sigma(w_a)=0.04$ to 0.004.  

The BAO alone results are worse than WL alone.  To understand
why we also plot results for BAO in the limit of perfect
prior knowledge of the bias parameters.  In this limit the
galaxy survey is also sensitive to growth so we label the
dashed curve as BAO+g.  This (unrealistic) case performs
better than WL.  This is as we expect since the galaxy power
spectra give us finer spatial and temporal sampling of the
dark matter power spectra than we get from weak lensing
with its broad kernels.  Likewise, we artificially remove all
growth information from WL by pretending that the gravitational
potentials are sourced by some unknown bias factors times the
dark matter density, factors that we then marginalize over.
We parameterize this b(z) in the same manner as we do with BAO.  
We see in the dot-dashed curve labeled WL-g that the WL
results are worse than BAO if we are unable to predict the
growth rate from our (non-bias) model parameters.  

An interesting feature of the BAO result is its near-independence
of \phz{} parameters. Since the effect of \phz{}
uncertainties is more localized in redshift space for galaxy angular 
power spectra, BAO data allow for useful constraints on the 
\phz{} parameters and therefore the dark energy constraints are 
less sensitive to \phz{} priors (Zhan 2006, 
in preparation). By combining BAO and WL, one can achieve dark energy 
constraints that are robust to the dominant uncertainties of either 
probes: the galaxy bias for BAO and \phz{} distribution for WL. 

Note that the experiment-combining procedure used for the DETF
report assumes that systematic error parameters for each experiment
are independent.  Taking the \phz{} parameters to be the
same for WL and BAO is an important difference and makes our
BAO + WL combination significantly more powerful than in the
DETF report.  To take advantage of this synergy with real data,
the analysis will either have to be done so that the galaxies
used for their correlation properties and those used for their
shape properties are weighted in the same manner, or the differences
in the populations used for WL and BAO will have to be modeled.

SN can measure relative distances more accurately at low redshift than
at high redshift.  In contrast, the smaller volumes available at low
redshift make WL and BAO distance and growth constraints weaker toward
lower redshifts.  This complementarity leads to remarkable reductions of
the product $\sigma(w_p) \times \sigma(w_a)$ from combinations of SN
with BAO or WL, as shown in Fig.~\ref{fig:fom}.  The complementarity 
is reflected in the orientations
of the error ellipses of WL and SN in Fig.~\ref{fig:w0wa} as well.
Furthermore, WL and/or BAO provide a normalization of the $D(z)$ curve,
and a strong constraint on curvature as discussed in the previous
section.  Given the curvature constraints from WL/BAO, the SN
$w_0$--$w_a$ error ellipse is nearly the same as that for a flat
Universe. Hence, the combination SN+WL does not change appreciably
with the curvature prior.

\section{Conclusions}
\label{sec:discuss}

Precision measurements of the mean curvature are interesting in their
own right and also important for reaching conclusions about dark
energy, due to the dark energy-curvature degeneracy.  We found that
the WL and BAO datasets, as we modeled them, would be capable of
constraining $\Omega_{\rm K}$ at the $10^{-3}$ level.  This tight constraint
is highly robust to photometric redshift errors, much more so than is
the case for the dark energy parameters, especially for WL.  Further,
the tight constraint essentially breaks the dark energy-curvature
degeneracy.

There are many challenges that must be met so that real surveys can
achieve our forecasted parameter constraints.  We have addressed
one which has been given much attention recently: photometric 
redshift errors.  We find that, given our (70-parameter) 
photometric redshift error model, and conservative prior
information on the parameters of this model, the galaxy correlations
serve to control the parameters of the model well enough that
the combined WL + BAO constraints on curvature and dark energy
are very strong.  

\acknowledgments We thank A. Albrecht, G. Bernstein, W. Freedman,
J. Hewitt, W. Hu, A. Kosowsky, M. Seiffert, N. Suntzeff, J. Wacker and
M.  Wood-Vasey for useful conversations.  This work was supported in
part at UCD by NASA grant NAG5-11098 and NSF grant No. 0307961.

\bibliography{/work3/knox/bib/cmb4}

\begin{thebibliography}{42}
\expandafter\ifx\csname natexlab\endcsname\relax\def\natexlab#1{#1}\fi

\bibitem[{{Aldering}(2005)}]{aldering05}
{Aldering}, G. 2005, New Astronomy Review, 49, 46

\bibitem[{{Bernstein}(2005)}]{bernstein05}
{Bernstein}, G. 2005, ArXiv Astrophysics e-prints

\bibitem[{{Bernstein} \& {Jain}(2004)}]{bernstein04}
{Bernstein}, G. \& {Jain}, B. 2004, \apj, 600, 17

\bibitem[{{de Bernardis} {et~al.}(2000){de Bernardis}, {Ade}, {Bock}, {Bond},
  {Borrill}, {Boscaleri}, {Coble}, {Crill}, {De Gasperis}, {Farese},
  {Ferreira}, {Ganga}, {Giacometti}, {Hivon}, {Hristov}, {Iacoangeli}, {Jaffe},
  {Lange}, {Martinis}, {Masi}, {Mason}, {Mauskopf}, {Melchiorri}, {Miglio},
  {Montroy}, {Netterfield}, {Pascale}, {Piacentini}, {Pogosyan}, {Prunet},
  {Rao}, {Romeo}, {Ruhl}, {Scaramuzzi}, {Sforna}, \&
  {Vittorio}}]{debernardis00}
{de Bernardis}, P., {Ade}, P.~A.~R., {Bock}, J.~J., {Bond}, J.~R., {Borrill},
  J., {Boscaleri}, A., {Coble}, K., {Crill}, B.~P., {De Gasperis}, G.,
  {Farese}, P.~C., {Ferreira}, P.~G., {Ganga}, K., {Giacometti}, M., {Hivon},
  E., {Hristov}, V.~V., {Iacoangeli}, A., {Jaffe}, A.~H., {Lange}, A.~E.,
  {Martinis}, L., {Masi}, S., {Mason}, P.~V., {Mauskopf}, P.~D., {Melchiorri},
  A., {Miglio}, L., {Montroy}, T., {Netterfield}, C.~B., {Pascale}, E.,
  {Piacentini}, F., {Pogosyan}, D., {Prunet}, S., {Rao}, S., {Romeo}, G.,
  {Ruhl}, J.~E., {Scaramuzzi}, F., {Sforna}, D., \& {Vittorio}, N. 2000, \nat,
  404, 955

\bibitem[{{Dodelson} \& {Knox}(2000)}]{dodelson00}
{Dodelson}, S. \& {Knox}, L. 2000, Physical Review Letters, 84, 3523

\bibitem[{{Eisenstein} {et~al.}(2005){Eisenstein}, {Zehavi}, {Hogg},
  {Scoccimarro}, {Blanton}, {Nichol}, {Scranton}, {Seo}, {Tegmark}, {Zheng},
  {Anderson}, {Annis}, {Bahcall}, {Brinkmann}, {Burles}, {Castander},
  {Connolly}, {Csabai}, {Doi}, {Fukugita}, {Frieman}, {Glazebrook}, {Gunn},
  {Hendry}, {Hennessy}, {Ivezi{\'c}}, {Kent}, {Knapp}, {Lin}, {Loh}, {Lupton},
  {Margon}, {McKay}, {Meiksin}, {Munn}, {Pope}, {Richmond}, {Schlegel},
  {Schneider}, {Shimasaku}, {Stoughton}, {Strauss}, {SubbaRao}, {Szalay},
  {Szapudi}, {Tucker}, {Yanny}, \& {York}}]{eisenstein05}
{Eisenstein}, D.~J., {Zehavi}, I., {Hogg}, D.~W., {Scoccimarro}, R., {Blanton},
  M.~R., {Nichol}, R.~C., {Scranton}, R., {Seo}, H.-J., {Tegmark}, M., {Zheng},
  Z., {Anderson}, S.~F., {Annis}, J., {Bahcall}, N., {Brinkmann}, J., {Burles},
  S., {Castander}, F.~J., {Connolly}, A., {Csabai}, I., {Doi}, M., {Fukugita},
  M., {Frieman}, J.~A., {Glazebrook}, K., {Gunn}, J.~E., {Hendry}, J.~S.,
  {Hennessy}, G., {Ivezi{\'c}}, Z., {Kent}, S., {Knapp}, G.~R., {Lin}, H.,
  {Loh}, Y.-S., {Lupton}, R.~H., {Margon}, B., {McKay}, T.~A., {Meiksin}, A.,
  {Munn}, J.~A., {Pope}, A., {Richmond}, M.~W., {Schlegel}, D., {Schneider},
  D.~P., {Shimasaku}, K., {Stoughton}, C., {Strauss}, M.~A., {SubbaRao}, M.,
  {Szalay}, A.~S., {Szapudi}, I., {Tucker}, D.~L., {Yanny}, B., \& {York},
  D.~G. 2005, \apj, 633, 560

\bibitem[{{Freedman} {et~al.}(2001){Freedman}, {Madore}, {Gibson}, {Ferrarese},
  {Kelson}, {Sakai}, {Mould}, {Kennicutt}, {Ford}, {Graham}, {Huchra},
  {Hughes}, {Illingworth}, {Macri}, \& {Stetson}}]{freedman01}
{Freedman}, W.~L., {Madore}, B.~F., {Gibson}, B.~K., {Ferrarese}, L., {Kelson},
  D.~D., {Sakai}, S., {Mould}, J.~R., {Kennicutt}, R.~C., {Ford}, H.~C.,
  {Graham}, J.~A., {Huchra}, J.~P., {Hughes}, S.~M.~G., {Illingworth}, G.~D.,
  {Macri}, L.~M., \& {Stetson}, P.~B. 2001, \apj, 553, 47

\bibitem[{{Freivogel} {et~al.}(2006){Freivogel}, {Kleban}, {Rodr{\'{\i}}guez
  Mart{\'{\i}}nez}, \& {Susskind}}]{freivogel06}
{Freivogel}, B., {Kleban}, M., {Rodr{\'{\i}}guez Mart{\'{\i}}nez}, M., \&
  {Susskind}, L. 2006, Journal of High Energy Physics, 3, 39

\bibitem[{{Garriga} {et~al.}(1999){Garriga}, {Tanaka}, \&
  {Vilenkin}}]{garriga99}
{Garriga}, J., {Tanaka}, T., \& {Vilenkin}, A. 1999, \prd, 60, 023501

\bibitem[{{Hirata} \& {Seljak}(2004)}]{hirata04}
{Hirata}, C.~M. \& {Seljak}, U. 2004, \prd, 70, 063526

\bibitem[{{Holz}(1998)}]{holz98}
{Holz}, D.~E. 1998, \apjl, 506, L1

\bibitem[{{Hu}(2005)}]{hu05}
{Hu}, W. 2005, in ASP Conf. Ser. 339: Observing Dark Energy, ed. S.~C. {Wolff}
  \& T.~R. {Lauer}, 215--+

\bibitem[{{Hu} \& {Jain}(2003)}]{hu03a}
{Hu}, W. \& {Jain}, B. 2003, ArXiv Astrophysics e-prints

\bibitem[{{Hu} \& {Jain}(2004)}]{hu04b}
---. 2004, \prd, 70, 043009

\bibitem[{{Huterer} \& {Cooray}(2005)}]{huterer05}
{Huterer}, D. \& {Cooray}, A. 2005, \prd, 71, 023506

\bibitem[{{Huterer} {et~al.}(2006){Huterer}, {Takada}, {Bernstein}, \&
  {Jain}}]{huterer06}
{Huterer}, D., {Takada}, M., {Bernstein}, G., \& {Jain}, B. 2006, \mnras, 366,
  101

\bibitem[{{Ilbert} {et~al.}(2006){Ilbert}, {Arnouts}, {McCracken},
  {Bolzonella}, {Bertin}, {Le Fevre}, {Mellier}, {Zamorani}, {Pello}, {Iovino},
  {Tresse}, {Bottini}, {Garilli}, {Le Brun}, {Maccagni}, {Picat}, {Scaramella},
  {Scodeggio}, {Vettolani}, {Zanichelli}, {Adami}, {Bardelli}, {Cappi},
  {Charlot}, {Ciliegi}, {Contini}, {Cucciati}, {Foucaud}, {Franzetti},
  {Gavignaud}, {Guzzo}, {Marano}, {Marinoni}, {Mazure}, {Meneux}, {Merighi},
  {Paltani}, {Pollo}, {Pozzetti}, {Radovich}, {Zucca}, {Bondi}, {Bongiorno},
  {Busarello}, {De La Torre}, {Gregorini}, {Lamareille}, {Mathez}, {Merluzzi},
  {Ripepi}, {Rizzo}, \& {Vergani}}]{ilbert06}
{Ilbert}, O., {Arnouts}, S., {McCracken}, H.~J., {Bolzonella}, M., {Bertin},
  E., {Le Fevre}, O., {Mellier}, Y., {Zamorani}, G., {Pello}, R., {Iovino}, A.,
  {Tresse}, L., {Bottini}, D., {Garilli}, B., {Le Brun}, V., {Maccagni}, D.,
  {Picat}, J.~P., {Scaramella}, R., {Scodeggio}, M., {Vettolani}, G.,
  {Zanichelli}, A., {Adami}, C., {Bardelli}, S., {Cappi}, A., {Charlot}, S.,
  {Ciliegi}, P., {Contini}, T., {Cucciati}, O., {Foucaud}, S., {Franzetti}, P.,
  {Gavignaud}, I., {Guzzo}, L., {Marano}, B., {Marinoni}, C., {Mazure}, A.,
  {Meneux}, B., {Merighi}, R., {Paltani}, S., {Pollo}, A., {Pozzetti}, L.,
  {Radovich}, M., {Zucca}, E., {Bondi}, M., {Bongiorno}, A., {Busarello}, G.,
  {De La Torre}, S., {Gregorini}, L., {Lamareille}, F., {Mathez}, G.,
  {Merluzzi}, P., {Ripepi}, V., {Rizzo}, D., \& {Vergani}, D. 2006, ArXiv
  Astrophysics e-prints

\bibitem[{{Jungman} {et~al.}(1996){Jungman}, {Kamionkowski}, {Kosowsky}, \&
  {Spergel}}]{jungman96b}
{Jungman}, G., {Kamionkowski}, M., {Kosowsky}, A., \& {Spergel}, D.~N. 1996,
  Physical Review Letters, 76, 1007

\bibitem[{{Kaplinghat} {et~al.}(2003){Kaplinghat}, {Knox}, \&
  {Song}}]{kaplinghat03b}
{Kaplinghat}, M., {Knox}, L., \& {Song}, Y. 2003, ArXiv Astrophysics e-prints,
  3344

\bibitem[{{Kim} {et~al.}(2004){Kim}, {Linder}, {Miquel}, \& {Mostek}}]{kim04}
{Kim}, A.~G., {Linder}, E.~V., {Miquel}, R., \& {Mostek}, N. 2004, \mnras, 347,
  909

\bibitem[{{Knox}(2006)}]{knox06}
{Knox}, L. 2006, \prd, 73, 023503

\bibitem[{{Knox} {et~al.}(2005){Knox}, {Song}, \& {Tyson}}]{knox05b}
{Knox}, L., {Song}, Y.~., \& {Tyson}, J.~A. 2005, ArXiv Astrophysics e-prints

\bibitem[{{Linde} \& {Mezhlumian}(1995)}]{linde95}
{Linde}, A. \& {Mezhlumian}, A. 1995, \prd, 52, 6789

\bibitem[{{Linder}(2005)}]{linder05d}
{Linder}, E.~V. 2005, Astroparticle Physics, 24, 391

\bibitem[{{Linder}(2006)}]{linder06}
---. 2006, ArXiv Astrophysics e-prints

\bibitem[{{Ma} {et~al.}(2006){Ma}, {Hu}, \& {Huterer}}]{ma06}
{Ma}, Z., {Hu}, W., \& {Huterer}, D. 2006, \apj, 636, 21

\bibitem[{{Mandelbaum} {et~al.}(2006){Mandelbaum}, {Hirata}, {Ishak}, {Seljak},
  \& {Brinkmann}}]{mandelbaum06}
{Mandelbaum}, R., {Hirata}, C.~M., {Ishak}, M., {Seljak}, U., \& {Brinkmann},
  J. 2006, \mnras, 367, 611

\bibitem[{{Martin} \& {Albrecht}(2006)}]{martin06}
{Martin}, D. \& {Albrecht}, A. 2006, ArXiv Astrophysics e-prints

\bibitem[{{Miller} {et~al.}(1999){Miller}, {Caldwell}, {Devlin}, {Dorwart},
  {Herbig}, {Nolta}, {Page}, {Puchalla}, {Torbet}, \& {Tran}}]{miller99}
{Miller}, A.~D., {Caldwell}, R., {Devlin}, M.~J., {Dorwart}, W.~B., {Herbig},
  T., {Nolta}, M.~R., {Page}, L.~A., {Puchalla}, J., {Torbet}, E., \& {Tran},
  H.~T. 1999, \apjl, 524, L1

\bibitem[{{Padmanabhan} {et~al.}(2006){Padmanabhan}, {Schlegel}, {Seljak},
  {Makarov}, {Bahcall}, {Blanton}, {Brinkmann}, {Eisenstein}, {Finkbeiner},
  {Gunn}, {Hogg}, {Ivezic}, {Knapp}, {Loveday}, {Lupton}, {Nichol},
  {Schneider}, {Strauss}, {Tegmark}, \& {York}}]{padmanabhan06}
{Padmanabhan}, N., {Schlegel}, D.~J., {Seljak}, U., {Makarov}, A., {Bahcall},
  N.~A., {Blanton}, M.~R., {Brinkmann}, J., {Eisenstein}, D.~J., {Finkbeiner},
  D.~P., {Gunn}, J.~E., {Hogg}, D.~W., {Ivezic}, Z., {Knapp}, G.~R., {Loveday},
  J., {Lupton}, R.~H., {Nichol}, R.~C., {Schneider}, D.~P., {Strauss}, M.~A.,
  {Tegmark}, M., \& {York}, D.~G. 2006, ArXiv Astrophysics e-prints

\bibitem[{{Phillips}(1993)}]{phillips93}
{Phillips}, M.~M. 1993, \apjl, 413, L105

\bibitem[{{Riess} \& {Livio}(2006)}]{riess06}
{Riess}, A.~G. \& {Livio}, M. 2006, ArXiv Astrophysics e-prints

\bibitem[{{Seo} \& {Eisenstein}(2003)}]{seo03}
{Seo}, H. \& {Eisenstein}, D.~J. 2003, \apj, 598, 720

\bibitem[{{Song} \& {Knox}(2004)}]{song04}
{Song}, Y. \& {Knox}, L. 2004, \prd, 70, 063510

\bibitem[{{Spergel} {et~al.}(2006){Spergel}, {Bean}, {Dore'}, {Nolta},
  {Bennett}, {Hinshaw}, {Jarosik}, {Komatsu}, {Page}, {Peiris}, {Verde},
  {Barnes}, {Halpern}, {Hill}, {Kogut}, {Limon}, {Meyer}, {Odegard}, {Tucker},
  {Weiland}, {Wollack}, \& {Wright}}]{spergel06}
{Spergel}, D.~N., {Bean}, R., {Dore'}, O., {Nolta}, M.~R., {Bennett}, C.~L.,
  {Hinshaw}, G., {Jarosik}, N., {Komatsu}, E., {Page}, L., {Peiris}, H.~V.,
  {Verde}, L., {Barnes}, C., {Halpern}, M., {Hill}, R.~S., {Kogut}, A.,
  {Limon}, M., {Meyer}, S.~S., {Odegard}, N., {Tucker}, G.~S., {Weiland},
  J.~L., {Wollack}, E., \& {Wright}, E.~L. 2006, ArXiv Astrophysics e-prints

\bibitem[{{Spergel} {et~al.}(2003){Spergel}, {Verde}, {Peiris}, {Komatsu},
  {Nolta}, {Bennett}, {Halpern}, {Hinshaw}, {Jarosik}, {Kogut}, {Limon},
  {Meyer}, {Page}, {Tucker}, {Weiland}, {Wollack}, \& {Wright}}]{spergel03}
{Spergel}, D.~N., {Verde}, L., {Peiris}, H.~V., {Komatsu}, E., {Nolta}, M.~R.,
  {Bennett}, C.~L., {Halpern}, M., {Hinshaw}, G., {Jarosik}, N., {Kogut}, A.,
  {Limon}, M., {Meyer}, S.~S., {Page}, L., {Tucker}, G.~S., {Weiland}, J.~L.,
  {Wollack}, E., \& {Wright}, E.~L. 2003, \apjs, 148, 175

\bibitem[{{Trotta}(2003)}]{trotta03}
{Trotta}, R. 2003, New Astronomy Review, 47, 769

\bibitem[{{Trotta} \& {Durrer}(2004)}]{trotta04}
{Trotta}, R. \& {Durrer}, R. 2004, ArXiv Astrophysics e-prints

\bibitem[{{Wittman}(2005)}]{wittman05}
{Wittman}, D. 2005, \apjl, 632, L5

\bibitem[{{Zhan} \& {Knox}(2005)}]{zhan06c}
{Zhan}, H. \& {Knox}, L. 2005, ArXiv Astrophysics e-prints

\bibitem[{{Zhan} {et~al.}(2006){Zhan}, {Knox}, {Tyson}, \&
  {Margoniner}}]{zhan06d}
{Zhan}, H., {Knox}, L., {Tyson}, J.~A., \& {Margoniner}, V. 2006, \apj, 640, 8

\bibitem[{{Zhang} {et~al.}(2005){Zhang}, {Hui}, \& {Stebbins}}]{zhang05}
{Zhang}, J., {Hui}, L., \& {Stebbins}, A. 2005, \apj, 635, 806

\end{thebibliography}
%\bibliography{cmb4}

\end{document}